\documentclass[a4paper]{article}
\usepackage[T1]{fontenc}
\usepackage[utf8]{inputenc}
\usepackage{amsmath}
\usepackage{amsthm}
\usepackage{esint}

\makeatletter

\newcommand*\LyXZeroWidthSpace{\hspace{0pt}}

\@ifundefined{date}{}{\date{}}
\usepackage[english]{babel}

\usepackage{color}
\usepackage{amsfonts}\usepackage{amsthm}\usepackage{amscd}\numberwithin{equation}{section}
\usepackage{bm}
\usepackage[all]{xy}

\usepackage{algorithm}
\usepackage{algorithmicx}
\usepackage{algpseudocode}
\usepackage[most]{tcolorbox}
\tcbuselibrary{listingsutf8}
\usepackage{float}
\tcbset{
  highlight algorithm/.style={
    colback=gray!10,
    colframe=black,
    boxrule=0.5pt,
    arc=2pt,
    left=6pt,
    right=6pt,
    top=6pt,
    bottom=6pt,
    width=\textwidth,
    sharp corners,
  }
}

\makeatother

\begin{document}
\begin{flushleft}
\textbf{OUJ-FTC-}23\\
 \medskip{}
 
\par\end{flushleft}

\begin{center}
{\Large A Gauge-Theoretic Formulation of Nambu Non-equilibrium Thermodynamics}{\Large\par}
\par\end{center}

\begin{center}
 
\par\end{center}

\begin{center}
\vspace{12pt}
 
\par\end{center}

\begin{center}
So Katagiri\textsuperscript{{*}}\footnote{So.Katagiri@gmail.com} 
\par\end{center}

\begin{center}
\textit{$^{*}$}{\small\textit{Major in Complex Systems Science, Graduate
School of Science and Engineering, Ibaraki University, Nakanarusawa-cho,
Hitachi, 316-8511, Japan.}}{\small\par}
\par\end{center}

\begin{center}
 
\par\end{center}

\begin{center}
\vspace{16pt}
 
\par\end{center}
\begin{abstract}
We present a gauge-theoretic formulation of Nambu non-equilibrium
thermodynamics (NNET) within a common gauge-fixed worldline framework.
In this formulation, the generalized thermodynamic force is described
by an Abelian thermodynamic connection, and ordinary Onsager dynamics
is recovered by fixing its temporal component through the dissipation
function. 

To extend this construction to NNET, we introduce a closed two-form
as additional geometric input. This form determines a divergence-free
Nambu vector field and thereby defines a characteristic direction
on the extended thermodynamic state space. The Onsager gauge condition
is then generalized so that the dissipative gauge fixing is imposed
along this Nambu characteristic direction. The resulting constitutive
relation reproduces the basic NNET evolution equation, combining nondissipative
Nambu circulation with dissipative Onsager relaxation. 

We further show that the localized BF-type expression is not an independent
pre-gauge-fixed action, but a differential-form representation of
the Nambu-adapted gauge-fixed worldline theory. The Faddeev--Popov
operator associated with the BRST gauge fixing becomes, on shell, the
negative of the material derivative along the full NNET flow, and its
time-independent zero-mode profiles satisfy the same equation
as first integrals of the deterministic dynamics. Finite residual
transformations obey a Hamilton--Jacobi equation governed by the same
Onsager tensor, implying that the gauge parameter is nonincreasing
along the original NNET flow. These results provide a unified geometric
interpretation of Onsager dissipation and Nambu dynamics within a
single gauge-fixing framework.
\end{abstract}

\section{Introduction}

Non-equilibrium thermodynamics was first formulated systematically
by Onsager as a linear-response theory based on thermodynamic forces,
fluxes, and reciprocal relations~\cite{Onsager_1931,Onsager_1931_2}.
This framework was subsequently developed by Onsager, Machlup, and
Hashitsume into variational and path-probability formulations that
incorporate fluctuations~\cite{Onsager_1953,Hashitsume_1952,Hashitsume_1956}.
In the present paper, we refer to this class of formulations collectively
as the Onsager--Machlup--Hashitsume (OMH) framework.

\LyXZeroWidthSpace{}

The scope of non-equilibrium thermodynamics was later broadened substantially
by the work of Prigogine and collaborators, who emphasized the distinction
between entropy production and entropy flux in open systems and developed
the theory of dissipative structures~\cite{Glansdorff_1964,GlansdorffPrigogine1971}.
More recent geometric approaches, including GENERIC, metriplectic
dynamics, and contact-geometric formulations, have provided powerful
frameworks for combining reversible and irreversible dynamics while
maintaining thermodynamic consistency~\cite{Kaufman_1984,Morrison_1984,Morrison_1986,grmela1997dynamics,ottinger1997dynamics,Ottinger_2005_Beyond,Bravetti_2017}.

\LyXZeroWidthSpace{}

Far from equilibrium, however, macroscopic dynamics may contain not
only monotonic relaxation but also persistent circulation, periodic
motion, spikes, bifurcations, and chaotic behavior. Motivated by this
observation, we have developed Nambu non-equilibrium thermodynamics
(NNET), in which the macroscopic velocity field is decomposed into
a nondissipative Nambu sector and a dissipative gradient sector~\cite{katagiri2022fluctuating,katagiri2025nambu1,katagiri2025nambu2}.
The Nambu bracket, originally introduced by Nambu as an $n$-ary generalization
of the Poisson bracket~\cite{Nambu_1973}, provides a natural description
of volume-preserving dynamics generated by several Hamiltonians. Its
geometric formulation was further developed by Takhtajan and others~\cite{Takhtajan_1994}.

\LyXZeroWidthSpace{}

In NNET, $S_{\mathrm{NB}}$ denotes the scalar potential generating
the dissipative sector. Throughout
this paper, we distinguish this potential from the physical thermodynamic
entropy unless additional thermodynamic assumptions justify their
identification. Rather, it is a state-space dissipation potential
generating the gradient part of the dynamics. In appropriate thermodynamic
settings it may acquire a more specific interpretation, while in the
general formulation its primary role is geometric. Previous work has
developed the axiomatic structure of NNET, its reduction properties,
and applications to explicit nonequilibrium systems~\cite{katagiri2025nambu1,katagiri2025nambu3,katagiri2025nambu2,NNET-Piston}.

\LyXZeroWidthSpace{}

The purpose of the present paper is not to reformulate NNET axiomatically,
but to clarify its relation to the gauge-fixing formulation of non-equilibrium
thermodynamics proposed previously~\cite{10.1093/ptep/pty102}. In
that formulation, the generalized thermodynamic force is represented
by a one-form

\begin{equation}
A=A_{i}da^{i},
\end{equation}

which, in a local integrable potential sector, satisfies

\begin{equation}
A=d_{a}S_{\mathrm{NB}},\qquad F=d_{a}A=0.
\end{equation}

The condition $F=0$ expresses local integrability of the state-space
one-form and should not be identified with reversibility. Dissipative
Onsager evolution can occur even when $A$ is exact. More generally,
a non-vanishing curvature $F\neq0$ represents non-integrability of
the generalized thermodynamic connection and may describe path-dependent
or non-potential thermodynamic forces.

\LyXZeroWidthSpace{}

To formulate the gauge structure dynamically, we extend the thermodynamic
state space by time and introduce the connection

\begin{equation}
\mathcal{A}=A_{i}da^{i}-A_{0}dt.
\end{equation}

The corresponding pre-gauge-fixed worldline action possesses the Abelian
redundancy

\begin{equation}
\mathcal{A}\rightarrow\mathcal{A}+d\epsilon
\end{equation}

up to endpoint terms. Onsager dynamics is obtained by imposing the
gauge condition

\begin{equation}
-\mathcal{A}(\partial_{t})=A_{0}=\Psi(A),\qquad\Psi(A)=\frac{1}{2}L^{ij}A_{i}A_{j},
\end{equation}

which yields the constitutive equation

\begin{equation}
\dot{a}^{i}=L^{ij}A_{j}.
\end{equation}

The main step of the present work is to extend this construction to
NNET. For a three-dimensional thermodynamic state space with volume
form $\Omega$, we introduce a closed two-form $B$, treated as additional
geometric input, and define the associated divergence-free Nambu vector
field $v_{H}$ through

\begin{equation}
\iota_{v_{H}}\Omega=B.
\end{equation}

In a regular local chart,

\begin{equation}
B=dH_{1}\wedge dH_{2},
\end{equation}

so that
\begin{equation}
v_{H}^{i}=\{a^{i},H_{1},H_{2}\}.
\end{equation}

The Nambu geometry therefore determines the characteristic direction

\begin{equation}
D_{H}=\partial_{t}+v_{H}
\end{equation}

on the extended thermodynamic space. We then generalize the Onsager
gauge condition from the direction $\partial_{t}$ to the Nambu characteristic
direction, 
\begin{equation}
-\mathcal{A}(D_{H})=\Psi(A).
\end{equation}

This Nambu-adapted gauge fixing gives
\begin{equation}
A_{0}=A_{i}v_{H}^{i}+\frac{1}{2}L^{ij}A_{i}A_{j},
\end{equation}

and the corresponding dual constitutive variation yields

\begin{equation}
\dot{a}^{i}=v_{H}^{i}+L^{ij}A_{j}.
\end{equation}

In the local potential sector $A_{i}=\partial_{i}S_{\mathrm{NB}}$, this becomes

\begin{equation}
\dot{a}^{i}=\{a^{i},H_{1},H_{2}\}+L^{ij}\partial_{j}S_{\mathrm{NB}},
\end{equation}

which is the basic NNET evolution equation.

\LyXZeroWidthSpace{}

The closed form $B$ is not derived from the Abelian gauge redundancy
itself. Rather, it specifies the nondissipative Nambu geometry independently,
while the gauge-fixing construction determines how this geometry is
combined with the Onsager dissipative sector. The localized BF/Chern--Simons-like
expression considered below should likewise not be interpreted as
a second independent pre-gauge-fixed action. Instead, it provides
a differential-form representation of the Nambu-adapted gauge-fixed
worldline theory. In this sense, the present formulation clarifies
the relation between the gauge-fixing description and the BF-type
representation while avoiding the mixing of worldline and bulk integration
domains.

\LyXZeroWidthSpace{}

The new content of the present work is therefore not the NNET evolution
equation by itself. Rather, it consists of a common gauge-fixed origin
for the Onsager and NNET constitutive relations, the identification
of the Nambu characteristic direction as the direction entering the
generalized gauge condition, the equivalence between the localized
BF-type representation and the Nambu-adapted gauge-fixed worldline
action, and the associated BRST/Faddeev--Popov structure. In particular,
the Faddeev--Popov operator of the Nambu-adapted gauge becomes, on shell,
the negative of the material derivative along the full NNET flow. Its
time-independent zero-mode profiles satisfy the same equation as first
integrals of the deterministic dynamics. Beyond the infinitesimal
level, we derive a nonlinear Hamilton--Jacobi equation for finite
residual transformations. The same transport tensor $L^{ij}$ determines
the quadratic term in this equation and the rate of change of the
residual parameter along the original NNET flow.

\LyXZeroWidthSpace{}

The organization of this paper is as follows. In Sec.~2, we review
the thermodynamic gauge theory and formulate the Onsager constitutive
relation in terms of an extended worldline connection and its gauge
fixing. In Sec.~3, we introduce the closed two-form $B$, construct
the Nambu characteristic direction, derive the Nambu-adapted gauge
condition, and show the equivalence between the worldline and localized
BF-type representations. In Sec.~4, we discuss the geometric meanings
of integrability, curvature, the dissipation potential $S_{\mathrm{NB}}$,
and the Nambu-adapted gauge. In Sec.~5, we formulate the associated
BRST structure and discuss the Faddeev--Popov operator, its zero
modes, and finite residual transformations. Finally, Sec.~6 summarizes
the scope, limitations, and future
directions of the present construction.

\section{Thermodynamic Gauge Theory}

In this section, we review the framework of non-equilibrium thermodynamics
as gauge fixing developed in our previous work~\cite{10.1093/ptep/pty102}.

We begin with the integrable potential sector of the thermodynamic
gauge formulation. In this sector, the thermodynamic one-form can
be written locally as

\begin{equation}
A=d_{a}S_{\mathrm{NB}}=A_{i}da^{i},
\end{equation}

where $a^{i}$ $(i=1,\dots,n)$ are coordinates on the thermodynamic
state space, $S_{\mathrm{NB}}$ is a scalar dissipation potential, and $A_{i}=\partial_{i}S_{\mathrm{NB}}$
is the generalized thermodynamic force conjugate to $a^{i}$. In this
integrable sector, one has

\begin{equation}
A_{i}=\frac{\partial S_{\mathrm{NB}}}{\partial a^{i}}.
\end{equation}

According to Stokes' theorem,

\begin{equation}
\oint_{\partial\Sigma}dS_{\mathrm{NB}}=\int_{\Sigma}ddS_{\mathrm{NB}}=0.
\end{equation}

This equation expresses the path-independence of the dissipation potential
as a state function. Equivalently, the thermodynamic one-form is exact,
and therefore its curvature vanishes.

Now consider a local translation of the dissipation potential,

\begin{equation}
\delta S_{\mathrm{NB}}=\epsilon.
\end{equation}

Under this transformation, the thermodynamic potential $A_{i}=\partial_{i}S_{\mathrm{NB}}$
changes as

\begin{equation}
\delta A=d\epsilon.
\end{equation}

Thus, $A_{i}$ behaves as an Abelian gauge potential associated with
local shifts of the dissipation potential. In the integrable thermodynamic
sector, the corresponding curvature is

\begin{equation}
F=dA,
\end{equation}

and since $A=dS_{\mathrm{NB}}$, one obtains

\begin{equation}
F=dA=d^{2}S_{\mathrm{NB}}=0.
\end{equation}

It is important to emphasize that the condition $F=0$ should not
be identified with reversibility itself. Rather, it expresses the
integrability of the thermodynamic one-form. In ordinary Onsager theory,
even irreversible relaxation processes may be described by a dissipation
potential $S_{\mathrm{NB}}(a)$, so that $A_{i}=\partial_{i}S_{\mathrm{NB}}$ and hence
$dA=0$ on the thermodynamic state space. The irreversibility of such
processes is encoded not by the curvature $dA$, but by the direction
of the dynamical flow and positivity of the production rate of the
dissipation potential $S_{\mathrm{NB}}$,

\begin{equation}
\dot{S}_{\mathrm{NB}}=A_{i}\dot{a}^{i}\geq0.
\end{equation}

Therefore, the flatness condition

\begin{equation}
F=dA=0
\end{equation}

should be understood as the condition that the thermodynamic connection
is locally pure gauge,

\begin{equation}
A=dS_{\mathrm{NB}},
\end{equation}

or, equivalently, that the dissipation potential is a locally well-defined
state function.

In the gauge-theoretic extension considered here, however, $A$ is
promoted to a more general thermodynamic connection, which need not
be globally or locally exact. In such a generalized setting,

\begin{equation}
F=dA\neq0
\end{equation}

measures the non-integrability of the thermodynamic connection. This
curvature should therefore be interpreted not as a necessary condition
for irreversibility, but as a geometrical obstruction to representing
the thermodynamic force globally as the gradient of a single dissipation
potential. It may describe non-potential generalized forces, path
dependence, or transverse/circulatory response associated with non-exact
thermodynamic cycles.

We now recall the deterministic Onsager variational principle~\cite{Onsager_1931,Onsager_1931_2}.
We distinguish the state-space dissipation potential $S_{\mathrm{NB}}$ from
the Onsager--Machlup--Hashitsume path functional $S_{\mathrm{OMH}}$.
The former generates the dissipative drift on the thermodynamic state
space, whereas the latter governs stochastic path weights. In the
deterministic Onsager sector, it is sufficient to use the dual dissipation
functions 
\begin{equation}
\Phi(\dot{a})=\frac{1}{2}L_{ij}\dot{a}^{i}\dot{a}^{j},\qquad\Psi(A)=\frac{1}{2}L^{ij}A_{i}A_{j}.
\end{equation}

Here, $L^{ij}$ is a symmetric, positive-definite transport tensor,
and $L_{ij}$ is its inverse:

\begin{equation}
L^{ij}L_{jk}=\delta_{k}^{i}.
\end{equation}

Varying the dual Onsager functional with respect to $A_{i}$,

\begin{equation}
\delta_{A_{i}}\left(A_{i}\dot{a}^{i}-\Psi(A)\right)=0,
\end{equation}

gives
\begin{equation}
\dot{a}^{i}-L^{ij}A_{j}=0.
\end{equation}

\subsection{Extended thermodynamic connection and gauge fixing}

To formulate the variational principle geometrically, we introduce
the extended thermodynamic state space

\begin{equation}
\tilde{\mathcal{M}}=\mathcal{M}_{\mathrm{th}}\times I_{t},
\end{equation}

with local coordinates $(a^{1},\dots,a^{n},t)$. On this space we
define the extended thermodynamic connection

\begin{equation}
\mathcal{A}=A_{i}da^{i}-A_{0}dt.
\end{equation}

The thermodynamic trajectory is defined by
\begin{equation}
\gamma:t\mapsto(a^{i}(t),t).
\end{equation}

The pullback of $\mathcal{A}$ along $\gamma$ is

\begin{equation}
\gamma^{*}\mathcal{A}=(A_{i}\dot{a}^{i}-A_{0})dt.
\end{equation}

The corresponding worldline action is therefore
\begin{equation}
I_{\mathrm{O}}[a,\mathcal{A}]=\int_{\gamma}\mathcal{A}=\int dt\left(A_{i}\dot{a}^{i}-A_{0}\right).
\end{equation}

The extended connection has the Abelian gauge redundancy

\begin{equation}
\mathcal{A}\to\mathcal{A}+d\epsilon,
\end{equation}

which in components reads
\begin{equation}
A_{i}\to A_{i}+\partial_{i}\epsilon,
\end{equation}

\begin{equation}
A_{0}\to A_{0}-\partial_{t}\epsilon.
\end{equation}

Under this transformation, the worldline action changes as
\begin{equation}
I_{\mathrm{O}}\to I_{\mathrm{O}}+\int_{\gamma}d\epsilon=I_{\mathrm{O}}+\epsilon_{\mathrm{f}}-\epsilon_{\mathrm{i}}.
\end{equation}

Under a gauge transformation, the action changes by the endpoint term
$\epsilon_{\mathrm{f}}-\epsilon_{\mathrm{i}}$. The fixed-endpoint variational problem
is therefore gauge invariant. The action itself is invariant when
$\epsilon_{\mathrm{f}}=\epsilon_{\mathrm{i}}$, in particular when $\epsilon$ vanishes
at both endpoints.

We now impose the Onsager gauge-fixing condition
\begin{equation}
A_{0}=\Psi(A)=\frac{1}{2}L^{ij}A_{i}A_{j}.
\end{equation}

Thus, 
\begin{equation}
I_{\mathrm{O}}^{\mathrm{gf}}=\int dt\left(A_{i}\dot{a}^{i}-\frac{1}{2}L^{ij}A_{i}A_{j}\right).
\end{equation}

Variation with respect to $A_{i}$ gives

\begin{equation}
\delta_{A_{i}}I_{\mathrm{O}}^{\mathrm{gf}}=\int dt\delta A_{i}\left(\dot{a}^{i}-L^{ij}A_{j}\right).
\end{equation}

Then,

\begin{equation}
\dot{a}^{i}=L^{ij}A_{j}.
\end{equation}

Equivalently, the constitutive equation gives
\begin{equation}
A_{i}=L_{ij}\dot{a}^{j}.
\end{equation}

Substituting this relation back into the action gives
\begin{equation}
I_{\mathrm{O}}^{\mathrm{gf}}=\frac{1}{2}\int dtL_{ij}\dot{a}^{i}\dot{a}^{j}=\int dt\Phi(\dot{a}).
\end{equation}

This exhibits the duality between the $A_{i}$-representation and
the $\dot{a}^{i}$-representation of the Onsager variational principle.

\subsection{Worldline current representation}

The worldline integral above can equivalently be represented as an
integral over the extended thermodynamic state space. Let $J_{\gamma}$
denote the distributional Poincaré-dual current of the trajectory
$\gamma$, defined by

\begin{equation}
\int_{\gamma}\omega=\int_{\tilde{\mathcal{M}}}\omega\wedge J_{\gamma}
\end{equation}

for every smooth one-form $\omega$ with suitable support.

The action can therefore be written as
\begin{equation}
I_{\mathrm{O}}=\int_{\tilde{\mathcal{M}}}\mathcal{A}\wedge J_{\gamma}.
\end{equation}

In local coordinates one may represent $J_{\gamma}$ distributionally
by

\begin{equation}
J_{\gamma}\propto\delta^{(n)}(a-a(t))\bigwedge_{i=1}^{n}\left(da^{i}-\dot{a}^{i}dt\right)
\end{equation}

up to the orientation convention.

This representation will be used in Sec.~3 so that both the trajectory
contribution and the BF/Chern--Simons-like contribution are formulated
on the same extended thermodynamic state space.

Before gauge fixing, configurations related by $\mathcal{A}\to\mathcal{A}+d\epsilon$
represent the redundancy of the extended connection.
Here the gauge redundancy refers to the pre-gauge-fixed connection
and its worldline functional. It should not be interpreted as an
equivalence between fully specified thermodynamic models with different
force or velocity fields. The thermodynamic interpretation of $A_i$
is assigned only after the constitutive gauge, a representative, and
the relevant initial or boundary data have been specified.

The Onsager condition
\begin{equation}
A_{0}=\frac{1}{2}L^{ij}A_{i}A_{j}
\end{equation}

specifies a gauge slice, which can still admit residual transformations.
Equivalently, the gauge-fixing condition is 

\begin{equation}
\chi_{\mathrm{O}}=A_{0}-\frac{1}{2}L^{ij}A_{i}A_{j}=0.
\end{equation}

Under an infinitesimal gauge transformation

\begin{equation}
\delta A_{i}=\partial_{i}\epsilon,\ \delta A_{0}=-\partial_{t}\epsilon,
\end{equation}

for constant $L^{ij}$, we obtain
\begin{equation}
\delta\chi_{\mathrm{O}}=-\partial_{t}\epsilon-L^{ij}A_{j}\partial_{i}\epsilon.
\end{equation}

Therefore, the infinitesimal residual gauge transformations preserving
the Onsager gauge satisfy
\begin{equation}
(\partial_{t}+L^{ij}A_{j}\partial_{i})\epsilon=0.
\end{equation}

Using the Onsager constitutive relation,
\begin{equation}
\dot{a}^{i}=L^{ij}A_{j},
\end{equation}

the residual-gauge condition becomes
\begin{equation}
(\partial_{t}+\dot{a}^{i}\partial_{i})\epsilon=0.
\end{equation}

A generic gauge transformation does not preserve this gauge condition;
therefore the gauge-fixed constitutive equation is not expected to
be invariant under an arbitrary transformation that moves the system
away from the chosen gauge. The finite residual condition and its
Hamilton--Jacobi structure are derived in Sec.~\ref{sec:finite_residual};
setting $v_H=0$ there gives the Onsager case.

For an autonomous representative $S_{\mathrm{NB}}(a)$, the rate of
increase of the dissipation potential is then

\begin{equation}
\frac{dS_{\mathrm{NB}}}{dt}=\frac{\partial S_{\mathrm{NB}}}{\partial a^{i}}\dot{a}^{i}=L^{ij}\frac{\partial S_{\mathrm{NB}}}{\partial a^{i}}\frac{\partial S_{\mathrm{NB}}}{\partial a^{j}}\geq0.
\end{equation}

Equivalently, using $A_{i}=\partial_{i}S_{\mathrm{NB}}$, this can be written
as

\begin{equation}
\frac{dS_{\mathrm{NB}}}{dt}=L^{ij}A_{i}A_{j}.
\end{equation}

This equation shows that the irreversibility of the Onsager process
is encoded in the dissipative flow and the positive-definiteness of
$L^{ij}$, rather than in a nonzero curvature $dA$ on the thermodynamic
state space.

In the gauge-fixed formulation, however, one may consider the extended
connection

\begin{equation}
\mathcal{A}=A_{i}da^{i}-A_{0}dt,\qquad A_{0}=\Psi(A).
\end{equation}

Even when the spatial part $A_{i}da^{i}$ is locally exact, the extended
connection can have nontrivial mixed components involving the time
direction. Indeed,

\begin{equation}
\mathcal{F}=d\mathcal{A}=d_{a}A+dt\wedge\left(\partial_{t}A+d_{a}A_{0}\right),
\end{equation}

where $d_{a}$ denotes the exterior derivative on the thermodynamic
state space. In the time-independent integrable Onsager sector, $d_{a}A=0$
and $\partial_{t}A=0$, so that

\begin{equation}
dt\wedge d_{a}A_{0}=dt\wedge d_{a}\Psi.
\end{equation}

For constant $L^{ij}$, the dissipation function $\Psi=\frac{1}{2}L^{ij}A_{i}A_{j}$
gives

\begin{equation}
\mathcal{F}=dt\wedge\partial_{k}\Psi\,da^{k}=dt\wedge L^{ij}(\partial_{k}A_{i})A_{j}\,da^{k}.
\end{equation}

If $A_{i}=\partial_{i}S_{\mathrm{NB}}$, this becomes

\begin{equation}
\mathcal{F}=dt\wedge L^{ij}\frac{\partial^{2}S_{\mathrm{NB}}}{\partial a^{i}\partial a^{k}}\frac{\partial S_{\mathrm{NB}}}{\partial a^{j}}\,da^{k}.
\end{equation}

Thus, in the gauge-fixed extended formulation, the curvature component
involving the time direction represents the state-space dependence
of the Onsager dissipation function $\Psi(A)$ along the thermodynamic
state space. It should not be confused with the statement that irreversibility
requires $d_{a}A\neq0$. Rather, it geometrically encodes how the
Onsager gauge-fixing condition changes along non-equilibrium trajectories.

The corresponding contribution over a two-dimensional surface $\Sigma$
in the extended thermodynamic space is

\begin{equation}
\int_{\Sigma}\mathcal{F}=\int_{\Sigma}L^{ij}\frac{\partial^{2}S_{\mathrm{NB}}}{\partial a^{i}\partial a^{k}}\frac{\partial S_{\mathrm{NB}}}{\partial a^{j}}\,dt\wedge da^{k}.
\end{equation}

This quantity should be interpreted as a geometric measure of the
variation of the dissipative structure in the extended thermodynamic
connection. The local production rate of the dissipation potential
itself is given by

\begin{equation}
\dot{S}_{\mathrm{NB}}=L^{ij}A_{i}A_{j}\geq0.
\end{equation}

As discussed above, the pre-gauge-fixed theory possesses the Abelian
redundancy.

This is an Abelian gauge redundancy associated with the thermodynamic
connection, rather than an internal Yang--Mills symmetry. The gauge-invariant
quantity is the curvature $F=dA$. Its vanishing on the thermodynamic
state space expresses the integrability of the thermodynamic one-form,
while its nonzero value in a generalized connection represents non-integrability
or the variation of the gauge-fixed dissipative structure.

In our previous work~\cite{10.1093/ptep/pty102}, we further extended
this framework by introducing a kinetic term for the thermodynamic
field $F$, analogous to that of the electromagnetic field. This nonlinear
extension allowed the description of damped periodic motions, thereby
providing a geometric formulation of relaxation phenomena in non-equilibrium
systems.

\section{Nambu-Adapted Gauge Fixing and the Emergence of NNET}

\subsection{Closed two-form and Nambu characteristic flow}

To extend the gauge-fixing formulation to Nambu non-equilibrium thermodynamics,
we work on a three-dimensional thermodynamic state space with volume
form

\begin{equation}
\Omega=da^{1}\wedge da^{2}\wedge da^{3}.
\end{equation}

Let $B$ be a closed two-form on this space: 
\begin{equation}
dB=0.
\end{equation}

A closed two-form $B$ determines locally a divergence-free vector
field $v_{H}$ through

\begin{equation}
\iota_{v_{H}}\Omega=B.
\end{equation}

Since $d\Omega=0$ and $dB=0$, Cartan's formula gives

\begin{equation}
\mathcal{L}_{v_{H}}\Omega=d(\iota_{v_{H}}\Omega)=dB=0,
\end{equation}

and hence $v_{H}$ is volume preserving.

In a regular local chart, we use the Clebsch representation

\begin{equation}
B=dH_{1}\wedge dH_{2},
\end{equation}

so that

\begin{equation}
v_{H}^{i}=\{a^{i},H_{1},H_{2}\}.
\end{equation}

We then define the corresponding characteristic direction on the extended
thermodynamic space by

\begin{equation}
D_{H}=\partial_{t}+v_{H}.
\end{equation}

This construction is motivated by the first-order geometric formulation
of Nambu mechanics due to Takhtajan~\cite{Takhtajan_1994}, in which
the Nambu trajectories arise as characteristic curves of the corresponding
differential-form structure. Here we use this observation only to
identify the distinguished Nambu characteristic direction $D_{H}$;
the closed form $B$ itself is treated as input geometric data.

\subsection{Nambu-adapted gauge fixing}

The Onsager gauge condition introduced in Sec.~2 can be written geometrically
as

\begin{equation}
-\mathcal{A}(\partial_{t})=\Psi(A).
\end{equation}

Once the Nambu structure is introduced, the direction entering the
gauge condition is generalized from $\partial_{t}$ to the Nambu characteristic
direction $D_{H}=\partial_{t}+v_{H}$. We therefore impose the Nambu-adapted
gauge condition

\begin{equation}
-\mathcal{A}(D_{H})=\Psi(A),
\end{equation}

or equivalently

\begin{equation}
A_{0}-A_{i}v_{H}^{i}-\frac{1}{2}L^{ij}A_{i}A_{j}=0.
\end{equation}

The Nambu structure is not derived from the gauge redundancy itself.
Rather, the closed form $B$, or equivalently the characteristic field
$v_{H}$, is supplied as additional geometric input. The role of the
present construction is to show how this nondissipative Nambu structure
can be incorporated consistently into the same gauge-fixed variational
framework as the Onsager dissipative sector.

\subsection{Localized BF-type representation}

We define the Onsager gauge-fixed connection by

\begin{equation}
\mathcal{A}_{\mathrm{O}}^{\mathrm{gf}}=A_{i}da^{i}-\Psi(A)dt.
\end{equation}

The localized BF-type representation is

\begin{equation}
I_{\mathrm{NNET}}^{\mathrm{loc}}=\int_{\tilde{\mathcal{M}}}\mathcal{A}_{\mathrm{O}}^{\mathrm{gf}}\wedge J_{\gamma}+\int_{\tilde{\mathcal{M}}}\rho_{\gamma}\mathcal{A}_{\mathrm{O}}^{\mathrm{gf}}\wedge B\wedge dt,\label{eq:locAction}
\end{equation}

where
\begin{equation}
\rho_{\gamma}(a,t)=\delta^{(3)}(a-a(t)).
\end{equation}

With the extended orientation $\widetilde{\Omega}=dt\wedge\Omega$
and the corresponding normalization of the Poincaré-dual current $J_{\gamma}$,
we have

\begin{equation}
\mathcal{A}_{\mathrm{O}}^{\mathrm{gf}}\wedge J_{\gamma}=\rho_{\gamma}(A_{i}\dot{a}^{i}-\Psi(A))\tilde{\Omega},
\end{equation}

\begin{equation}
\rho_{\gamma}\mathcal{A}_{\mathrm{O}}^{\mathrm{gf}}\wedge B\wedge dt=-\rho_{\gamma}A_{i}v_{H}^{i}\tilde{\Omega}.
\end{equation}

The second identity follows from $B=\iota_{v_{H}}\Omega$. Integrating
over the state-space coordinates therefore gives
\begin{equation}
I_{\mathrm{NNET}}^{\mathrm{loc}}=\int dt\left(A_{i}(\dot{a}^{i}-v_{H}^{i})-\frac{1}{2}L^{ij}A_{i}A_{j}\right).
\end{equation}

On the other hand, imposing the Nambu-adapted gauge condition
\begin{equation}
A_{0}=\Psi+A_{i}v_{H}^{i},
\end{equation}

and writing the worldline action before gauge fixing as $I_{\mathrm{pre}}\equiv\int_{\gamma}\mathcal{A}$,
we obtain

\begin{equation}
\left.I_{\mathrm{pre}}\right|_{\chi_{H}=0}=\int_{\gamma}\mathcal{A}_{H}^{\mathrm{gf}}=I_{\mathrm{NNET}}^{\mathrm{loc}},
\end{equation}

\begin{equation}
\mathcal{A}_{H}^{\mathrm{gf}}=A_{i}da^{i}-\left(\Psi(A)+A_{i}v_{H}^{i}\right)dt.
\end{equation}

Thus, the localized BF-type expression represents the Nambu-adapted
gauge-fixed worldline action; it is not an independent gauge-invariant
action before gauge fixing.

\subsection{Variational derivation of NNET}

We first derive the NNET constitutive equation from the gauge-fixed
dual functional

\begin{equation}
I_{\mathrm{NNET}}^{\mathrm{dual}}=\int dt\left(A_{i}\left(\dot{a}^{i}-v_{H}^{i}\right)-\frac{1}{2}L^{ij}A_{i}A_{j}\right).
\end{equation}

Varying $A_{i}$ while keeping the trajectory and the Nambu background
fixed gives
\begin{equation}
\delta_{A}I_{\mathrm{NNET}}^{\mathrm{dual}}=\int dt\delta A_{i}\left(\dot{a}^{i}-v_{H}^{i}-L^{ij}A_{j}\right).
\end{equation}

Requiring this variation to vanish for arbitrary $\delta A_{i}$ yields
\begin{equation}
\dot{a}^{i}=v_{H}^{i}+L^{ij}A_{j},
\end{equation}

or, equivalently,
\begin{equation}
\dot{a}^{i}=\{a^{i},H_{1},H_{2}\}+L^{ij}A_{j}.
\end{equation}

In the local potential sector, $A_{i}=\partial_{i}S_{\mathrm{NB}}$,
this reduces to the basic NNET evolution equation, 
\begin{equation}
\dot{a}^{i}=\{a^{i},H_{1},H_{2}\}+L^{ij}\partial_{j}S_{\mathrm{NB}}.
\end{equation}

This is a dual Onsager variation that determines the instantaneous
relation between the generalized thermodynamic force and the velocity
relative to the Nambu flow. Here, $A_{i}$ and $a^{i}$ are not treated
as an unconstrained canonical pair to be varied independently in a
Hamiltonian action.

Separately, the deterministic NNET flow admits a trajectory-level
characterization through the nonnegative residual functional
\begin{equation}
I_{\mathrm{path}}[a]=\frac{1}{2}\int dt\left(\dot{a}^{i}-f^{i}\right)L_{ij}\left(\dot{a}^{j}-f^{j}\right),\label{eq:path_residual}
\end{equation}

where

\begin{equation}
f^{i}(a)=v_{H}^{i}+L^{ij}\partial_{j}S_{\mathrm{NB}}.
\end{equation}

Since $L_{ij}$ is positive definite, $I_{\mathrm{path}}[a]\geq0$, with
\begin{equation}
I_{\mathrm{path}}[a]=0\Longleftrightarrow\dot{a}^{i}=f^{i}.
\end{equation}

Then,
\begin{equation}
\dot{a}^{i}=f^{i}=v_{H}^{i}+L^{ij}\partial_{j}S_{\mathrm{NB}}.
\end{equation}

For a prescribed initial state, any trajectory satisfying the NNET
evolution equation makes $I_{\mathrm{path}}$ vanish and therefore realizes
its global minimum. We use this non-negative residual functional as
the trajectory-level variational characterization of the deterministic
NNET flow.

The relation to the dual Onsager functional can be made explicit
by setting $w^i=\dot{a}^i-v_H^i$. The symmetry and positive
definiteness of $L^{ij}$ and the inverse relation between $L^{ij}$
and $L_{ij}$ give the identity
\[
\begin{aligned}
&\frac12 L_{ij}w^i w^j
 -\left(A_i w^i-\frac12 L^{ij}A_i A_j\right)\\
&\qquad=\frac12 L_{ij}
 \left(w^i-L^{ik}A_k\right)
 \left(w^j-L^{j\ell}A_\ell\right)\geq0.
\end{aligned}
\]
At the prescribed force $A_i=\partial_i S_{\mathrm{NB}}$, the right-hand
side is precisely the integrand in Eq.~\eqref{eq:path_residual}.
Thus $I_{\mathrm{path}}$ is the nonnegative gap associated with the same
Onsager duality, rather than the dual functional itself.

In a stochastic extension with covariance proportional to $L^{ij}$,
this quadratic residual functional is closely related to the quadratic
part of an Onsager--Machlup--Hashitsume path weight; a complete
stochastic formulation additionally requires the appropriate discretization
and Jacobian terms.

\subsection{Evolution of $S_{\mathrm{NB}}$ and $H_{a}$}

The balance identities in this subsection refer to autonomous
representatives, with no explicit time dependence in $S_{\mathrm{NB}}$
or $H_a$. The same convention applies to the balance identities
summarized in Sec.~4. For explicitly time-dependent representatives,
the corresponding partial time derivatives must be added to the
total-derivative formulas.

Using the NNET evolution equation and $\dot{S}_{\mathrm{NB}}=\partial_{i}S_{\mathrm{NB}}\dot{a}^{i}$,
we obtain
\begin{equation}
\dot{S}_{\mathrm{NB}}=\{S_{\mathrm{NB}},H_{1},H_{2}\}+L^{ij}\partial_{i}S_{\mathrm{NB}}\partial_{j}S_{\mathrm{NB}}.
\end{equation}

The contribution of the dissipative sector is
\begin{equation}
\sigma_{\mathrm{NB}}\equiv L^{ij}\partial_{i}S_{\mathrm{NB}}\partial_{j}S_{\mathrm{NB}}\geq0.
\end{equation}

In general, $\sigma_{\mathrm{NB}}$ should not be identified automatically
with the physical thermodynamic entropy production, since $S_{\mathrm{NB}}$
need not coincide with the thermodynamic entropy.

Similarly, the Nambu Hamiltonians $H_{a}$ ($a=1,2$)
evolve according to 
\begin{equation}
\dot{H}_{a}=L^{ij}\partial_{i}H_{a}\partial_{j}S_{\mathrm{NB}}.
\end{equation}

The Nambu contribution vanishes by complete antisymmetry.

In the special case in which $S_{\mathrm{NB}}$ is preserved by the Nambu flow,
\begin{equation}
v_{H}^{i}\partial_{i}S_{\mathrm{NB}}=0,
\end{equation}

which is equivalent to
\begin{equation}
\{S_{\mathrm{NB}},H_{1},H_{2}\}=0,
\end{equation}

we obtain
\begin{equation}
\dot{S}_{\mathrm{NB}}=L^{ij}\partial_{i}S_{\mathrm{NB}}\partial_{j}S_{\mathrm{NB}}\geq0.
\end{equation}

In this special case, $S_{\mathrm{NB}}$ is nondecreasing along NNET
trajectories.

\section{Geometric Interpretation of the Gauge-Fixed NNET Structure}

\subsection{Integrability, gauge fixing, and curvature}

The flatness condition of the state-space one-form,

\begin{equation}
F=d_{a}A=0,
\end{equation}

expresses local integrability. In the potential sector this implies

\begin{equation}
A=d_{a}S_{\mathrm{NB}}.
\end{equation}

This condition should not be identified with reversibility. Dissipative
Onsager evolution can occur even when the state-space connection is
exact.

\subsection{Extended curvature after gauge fixing}

Even when $F=d_{a}A=0$, the extended curvature
\begin{equation}
\mathcal{F}=d\mathcal{A}=F+dt\wedge\left(\partial_{t}A+d_{a}A_{0}\right)
\end{equation}

need not vanish. The mixed components involving $dt$ encode the variation
of the chosen gauge-fixed temporal component $A_{0}$ over the thermodynamic
state space. Thus, nonzero extended curvature should not be interpreted
as the origin of irreversibility itself.

\subsection{Geometric meaning of the Nambu-adapted gauge}

In the pure Onsager sector, the gauge condition fixes the temporal
component of the connection along the direction $\partial_{t}$. After
introducing the Nambu geometry, the distinguished direction becomes
the characteristic vector

\begin{equation}
D_{H}=\partial_{t}+v_{H}.
\end{equation}

The Nambu-adapted gauge condition

\begin{equation}
-\mathcal{A}(D_{H})=\Psi(A)
\end{equation}

therefore states that the dissipative gauge fixing is imposed along
the characteristic flow of the nondissipative Nambu sector.

\subsection{Role of the closed form $B$}

The closed form $B$ is not generated by the Abelian gauge redundancy.
It represents additional geometric data specifying the nondissipative
circulation. Through
\begin{equation}
\iota_{v_{H}}\Omega=B,
\end{equation}

it determines the volume-preserving Nambu vector field. The gauge-theoretic
construction then specifies how this independently supplied nondissipative
geometry is combined with the Onsager dissipative sector.

\subsection{The localized BF-type expression}

The localized BF-type expression should therefore be interpreted as
a geometric representation of the Nambu-adapted gauge-fixed worldline
theory, rather than as an independent second dynamical principle.
Its role is to encode the contraction $A_{i}v_{H}^{i}$ geometrically
through the closed form $B$.

\subsection{Physical meaning of $S_{\mathrm{NB}}$}

The scalar $S_{\mathrm{NB}}$ should in general be distinguished from
the thermodynamic entropy $S_{\mathrm{th}}$. It is the state-space potential
generating the dissipative sector. Consequently,
\begin{equation}
\sigma_{\mathrm{NB}}=L^{ij}\partial_{i}S_{\mathrm{NB}}\partial_{j}S_{\mathrm{NB}}
\end{equation}

represents the nonnegative production rate associated with the dissipative
potential. Its identification with physical entropy production requires
additional thermodynamic assumptions.

\subsection{Special monotonic case}

When

\begin{equation}
v_{H}^{i}\partial_{i}S_{\mathrm{NB}}=0,
\end{equation}

the Nambu flow is tangent to the level sets of $S_{\mathrm{NB}}$. The nondissipative
circulation therefore leaves $S_{\mathrm{NB}}$ unchanged, and its evolution
is entirely governed by the dissipative gradient sector:

\begin{equation}
\dot{S}_{\mathrm{NB}}=L^{ij}\partial_{i}S_{\mathrm{NB}}\partial_{j}S_{\mathrm{NB}}\geq0.
\end{equation}

In this special case, $S_{\mathrm{NB}}$ is nondecreasing along NNET
trajectories.

\LyXZeroWidthSpace{}

The geometric structure may therefore be summarized as a hierarchy.
The exact state-space one-form describes the integrable potential
sector. Onsager dissipation is introduced by fixing the temporal component
of the extended connection. NNET is obtained when this gauge condition
is imposed along the Nambu characteristic direction determined by
the additional closed form $B$.

\begin{equation}
\begin{aligned}{\rm Integrable}\ {\rm potential\ {\rm sector}} &  & A=d_{a}S_{\mathrm{NB}},F=0,\\
{\rm Onsager\ {\rm sector}} &  & -\mathcal{A}(\partial_{t})=\Psi(A),\\
{\rm NNET\ {\rm sector}} &  & -\mathcal{A}(D_{H})=\Psi(A),\ D_{H}=\partial_{t}+v_{H}.
\end{aligned}
\label{eq:hierarchy}
\end{equation}

$v_{H}$ is independently specified by the closed form $B$.

\section{BRST Structure}

\subsection{BRST transformations and the Onsager gauge}

As a further issue, let us examine the BRST structure that appears
after gauge fixing~\cite{BecchiRouetStora_1976,FaddeevPopov_1967}.
The BRST construction introduced below should not be regarded as an
independent justification for the thermodynamic gauge redundancy.
The Abelian redundancy has already been identified at the level of
the pre-gauge-fixed worldline action in Sec.~2. Here BRST symmetry
is used only as the algebraic implementation of that previously specified
redundancy and of the corresponding gauge-fixing conditions.

At the formal classical level, we implement the Abelian gauge redundancy
using a Grassmann-odd ghost field $C(a,t)$. In a local potential
chart, the BRST transformations are
\begin{equation}
s_{\mathrm{g}}S_{\mathrm{NB}}=C,\qquad s_{\mathrm{g}}A_{i}=\partial_{i}C,\qquad s_{\mathrm{g}}A_{0}=-\partial_{t}C,
\end{equation}
\begin{equation}
s_{\mathrm{g}}C=0.
\end{equation}

We further introduce the anti-ghost $\bar{C}$ and the Nakanishi--Lautrup
auxiliary field $b$, with

\begin{equation}
s_{\mathrm{g}}\bar{C}=b,\qquad s_{\mathrm{g}}b=0.
\end{equation}

Then the nilpotency condition

\begin{equation}
s_{\mathrm{g}}^{2}=0
\end{equation}

is satisfied.

For the Onsager-type gauge fixing, we define the gauge-fixing function

\begin{equation}
\chi_{\mathrm{O}}\equiv A_{0}-\frac{1}{2}L^{ij}A_{i}A_{j}.
\end{equation}

For constant $L^{ij}$, 
\begin{equation}
s_{\mathrm{g}}\chi_{\mathrm{O}}=-\partial_{t}C-L^{ij}A_{j}\partial_{i}C
\end{equation}

\begin{equation}
=-(\partial_{t}+L^{ij}A_{j}\partial_{i})C.
\end{equation}

Therefore
\begin{equation}
s_{\mathrm{g}}\chi_{\mathrm{O}}=-(\partial_{t}+v_{\mathrm{O}}^{i}\partial_{i})C,\qquad v_{\mathrm{O}}^{i}=L^{ij}A_{j}.
\end{equation}

On shell, 
\begin{equation}
v_{\mathrm{O}}^{i}=\dot{a}^{i},
\end{equation}

so 
\begin{equation}
s_{\mathrm{g}}\chi_{\mathrm{O}}=-(\partial_{t}+\dot{a}^{i}\partial_{i})C.
\end{equation}

The Faddeev--Popov operator obtained from the BRST variation of the
Onsager gauge condition is the negative of the differential operator that
appeared in Sec.~2 as the residual-gauge condition. Thus the BRST
ghost probes infinitesimal gauge transformations that preserve the
chosen Onsager gauge slice.

This condition identifies the time component $A_{0}$ with the dissipation
function

\begin{equation}
\Psi(A)=\frac{1}{2}L^{ij}A_{i}A_{j}.
\end{equation}

The corresponding gauge-fixing fermion is chosen as

\begin{equation}
\Psi_{\mathrm{gf}}^{\mathrm{O}}=\int dt\,\bar{C}\left(\chi_{\mathrm{O}}+\frac{\alpha}{2}b\right),
\end{equation}

where $\alpha$ is a gauge parameter. The gauge-fixing action is BRST
exact:

\begin{equation}
I_{\mathrm{gf}}^{\mathrm{O}}=s_{\mathrm{g}}\Psi_{\mathrm{gf}}^{\mathrm{O}}.
\end{equation}

Explicitly, one obtains

\begin{equation}
I_{\mathrm{gf}}^{\mathrm{O}}=\int dt\left[b\chi_{\mathrm{O}}+\frac{\alpha}{2}b^{2}-\bar{C}s_{\mathrm{g}}\chi_{\mathrm{O}}\right].
\end{equation}

Therefore, the gauge-ghost action is

\begin{equation}
I_{\mathrm{gh}}^{\mathrm{O}}=\int dt\,\bar{C}\left(\partial_{t}+L^{ij}A_{j}\partial_{i}\right)C.
\end{equation}

On shell, using the Onsager equation $\dot{a}^{i}=L^{ij}A_{j}$, this
becomes

\begin{equation}
I_{\mathrm{gh}}^{\mathrm{O}}=\int dt\,\bar{C}\left(\partial_{t}+\dot{a}^{i}\partial_{i}\right)C.
\end{equation}

Thus, the gauge ghost $C$ is transported as a scalar ghost along
the Onsager gradient flow.

\subsection{BRST structure of the Nambu-adapted gauge}
\label{sec:brst_nambu}

We now apply the same BRST construction to the Nambu-adapted gauge
condition introduced in Sec.~3,

\begin{equation}
\chi_{H}=A_{0}-A_{i}v_{H}^{i}-\frac{1}{2}L^{ij}A_{i}A_{j}.
\label{eq:chiH_brst}
\end{equation}

In the present work $B$ is treated as fixed background geometric
data, and no independent BRST symmetry is assigned to the $B$-sector.

In this case, $B$ and $v_{H}$ are input data and we set
\begin{equation}
s_{\mathrm{g}}B=0,\qquad s_{\mathrm{g}}v_{H}^{i}=0.
\end{equation}

We choose the gauge-fixing fermion
\begin{equation}
\Psi_{\mathrm{gf}}^{H}=\int dt\bar{C}\left(\chi_{H}+\frac{\alpha}{2}b\right),
\end{equation}
\begin{equation}
I_{\mathrm{gf}}^{H}=s_{\mathrm{g}}\Psi_{\mathrm{gf}}^{H}.
\end{equation}

Then,
\begin{equation}
\begin{aligned}s_{\mathrm{g}}\chi_{H} & =-\partial_{t}C-v_{H}^{i}\partial_{i}C-L^{ij}A_{j}\partial_{i}C\\
 & =-\left(\partial_{t}+\left(v_{H}^{i}+L^{ij}A_{j}\right)\partial_{i}\right)C.
\end{aligned}
\end{equation}

Defining the Faddeev--Popov operator by $s_{\mathrm{g}}\chi_{H}=M_{H}C$,
we obtain
\begin{equation}
M_{H}=-\partial_{t}-(v_{H}^{i}+L^{ij}A_{j})\partial_{i}.
\label{eq:fp_nambu}
\end{equation}

On shell, the NNET constitutive relation gives
$v_{H}^{i}+L^{ij}A_{j}=\dot{a}^{i}$. Thus, $M_{H}$ is the negative
of the material derivative along the full NNET flow:
\begin{equation}
M_{H}=-\left(\partial_{t}+\dot{a}^{i}\partial_{i}\right).
\end{equation}

\subsection{Ghost zero modes and first integrals}
\label{sec:ghost_integrals}

This result is useful because the zero modes of the gauge-fixing ghost
satisfy

\begin{equation}
M_{H}C=-\left(\partial_{t}+v^{i}\partial_{i}\right)C=0,\qquad v^{i}=v_{H}^{i}+L^{ij}A_{j}.
\end{equation}

For a time-independent ghost $C(a)$, this reduces to

\begin{equation}
v^{i}\partial_{i}C=0.
\end{equation}

Along a deterministic NNET trajectory, this gives

\begin{equation}
\frac{dC}{dt}=v^{i}\partial_{i}C=0.
\end{equation}

Thus, the scalar coefficient functions of a time-independent ghost
zero mode are constant along NNET trajectories and satisfy the same
equation as first integrals of the deterministic flow.

The statements above concern the local kernel of the differential
expression. The existence of admissible global zero modes additionally
depends on the chosen function space and boundary data.

\subsection{Finite residual transformations and Hamilton--Jacobi structure}
\label{sec:finite_residual}

We now retain the finite terms in the gauge transformation. Let
$\epsilon(a,t)$ be a real, smooth function on a local chart, and keep
$B$ and $L^{ij}$ fixed under the transformation. Write
\begin{equation}
A_i'=A_i+\partial_i\epsilon,\qquad
A_0'=A_0-\partial_t\epsilon,\qquad
f^i=v_H^i+L^{ij}A_j.
\label{eq:finite_gauge_data}
\end{equation}
Direct substitution into the gauge function in
Eq.~\eqref{eq:chiH_brst} gives the exact identity
\begin{equation}
\chi_H[\mathcal{A}']
=\chi_H[\mathcal{A}]
-\left(\partial_t+f^i\partial_i\right)\epsilon
-\frac12 L^{ij}\partial_i\epsilon\,\partial_j\epsilon.
\label{eq:finite_chi_change}
\end{equation}
Thus, if $\chi_H[\mathcal{A}]=0$, the transformed connection remains
on the same gauge slice if and only if
\begin{equation}
\partial_t\epsilon+f^i\partial_i\epsilon
+\frac12 L^{ij}\partial_i\epsilon\,\partial_j\epsilon=0.
\label{eq:finite_residual_hj}
\end{equation}
Here $\partial_t$ denotes differentiation at fixed thermodynamic
coordinates. Equation~\eqref{eq:finite_residual_hj} is a
Hamilton--Jacobi equation with Hamiltonian
\begin{equation}
\mathcal{H}_{\mathrm{res}}(a,p,t)
=f^i(a,t)p_i+\frac12 L^{ij}p_i p_j,
\qquad p_i=\partial_i\epsilon.
\label{eq:residual_hamiltonian}
\end{equation}
In particular, wherever the original constitutive drift vanishes,
$f^i=0$, it reduces to
\begin{equation}
\partial_t\epsilon
=-\frac12 L^{ij}\partial_i\epsilon\,\partial_j\epsilon.
\label{eq:residual_drift_free}
\end{equation}
If $f^i=0$ throughout a local region and $L^{ij}$ is constant, this
has the quadratic Hamilton--Jacobi form of a free particle. The
condition here is vanishing drift, not merely time independence of
the coefficients.

More generally, along a trajectory $a(t)$ of the original NNET drift,
$\dot a^i=f^i$, Eq.~\eqref{eq:finite_residual_hj} implies
\begin{equation}
\frac{d}{dt}\epsilon(a(t),t)
=\left(\partial_t+f^i\partial_i\right)\epsilon
=-\frac12 L^{ij}\partial_i\epsilon\,\partial_j\epsilon
\leq 0.
\label{eq:residual_monotonicity}
\end{equation}
The inequality follows from the positive definiteness of $L^{ij}$;
pointwise equality holds precisely when $\partial_i\epsilon=0$.
Thus finite residual shifts are nonincreasing along the original
constitutive flow, with a rate determined by the same Onsager
quadratic form that controls the dissipative sector.

In a local potential chart, the spatial part of the transformation
is $S_{\mathrm{NB}}'=S_{\mathrm{NB}}+\epsilon$, so the transformed
constitutive drift is
\begin{equation}
f'^i=v_H^i+L^{ij}A_j'
=f^i+L^{ij}\partial_j\epsilon.
\label{eq:residual_drift_change}
\end{equation}
The Hamilton--Jacobi characteristic velocity is
$\partial\mathcal{H}_{\mathrm{res}}/\partial p_i=f'^i$ at
$p_i=\partial_i\epsilon$. It must therefore be distinguished from
the original drift used in Eq.~\eqref{eq:residual_monotonicity}.
Indeed, with $D_{f'}=\partial_t+f'^i\partial_i$, the corresponding
rate has the opposite sign:
\[
D_{f'}\epsilon=\frac12 L^{ij}\partial_i\epsilon\,\partial_j\epsilon.
\]

To compare with the Faddeev--Popov description, consider a smooth
one-parameter family of residual shifts through the identity,
$\epsilon_\lambda=\lambda\eta+O(\lambda^2)$.
The term of first order in $\lambda$ in
Eq.~\eqref{eq:finite_residual_hj} gives
\begin{equation}
\left(\partial_t+f^i\partial_i\right)\eta=0
\quad\Longleftrightarrow\quad M_H\eta=0.
\label{eq:residual_linearization}
\end{equation}
Consequently, the transport equation for infinitesimal residual
parameters, and its time-independent first-integral sector, are
precisely the linearization of the finite Hamilton--Jacobi condition.
The quadratic term controls the departure from conservation at finite
amplitude.

There is also a direct connection to a familiar nonlinear evolution
equation. In a region where $f^i=0$ and $L^{ij}$ is constant, define
$u^i=L^{ij}\partial_j\epsilon$. Taking a spatial derivative of
Eq.~\eqref{eq:residual_drift_free} and using the symmetry of the
Hessian of $\epsilon$ gives
\begin{equation}
\partial_t u^i+u^j\partial_j u^i=0,
\label{eq:residual_burgers}
\end{equation}
which is the inviscid Burgers equation. These identities are local
statements on a domain where $\epsilon$ remains smooth.

The finite residual equation therefore constrains how the force and
dissipation-potential representatives can change while preserving the
same temporal gauge condition. Its monotonicity is a property of the
residual parameter; no identification of $\epsilon$ with physical
thermodynamic entropy is required. Since Eq.~\eqref{eq:residual_drift_change}
need not leave $f^i$ unchanged, preserving the gauge condition is
distinct from preserving a specified constitutive trajectory. A
particular constitutive model also specifies the representative and
its initial or boundary data, while the Hamilton--Jacobi equation
describes the constrained relation between representatives.

\section{Discussion and Outlook}

\subsection{Summary of results}

The central result of the present work is that the Onsager and NNET
constitutive equations can be formulated within a common gauge-fixed
worldline framework. The pre-gauge-fixed theory is defined by the
extended thermodynamic connection

\begin{equation}
\mathcal{A}=A_{i}da^{i}-A_{0}dt,
\end{equation}

whose worldline action is invariant under the Abelian transformation
$\mathcal{A}\to\mathcal{A}+d\epsilon$, up to endpoint terms. The
Onsager constitutive relation follows from the gauge condition

\begin{equation}
-\mathcal{A}(\partial_{t})=\Psi(A),
\end{equation}

whereas NNET is obtained by imposing the same dissipative gauge condition
along the Nambu characteristic direction

\begin{equation}
D_{H}=\partial_{t}+v_{H}.
\end{equation}

\subsection{Geometric input and derived structure}

An important limitation of the present construction is that the Nambu
structure itself is not generated by the Abelian gauge redundancy.
The closed form $B$, or equivalently the divergence-free vector field
$v_{H}$, is supplied as additional geometric input. What is derived
here is the way in which this independently specified nondissipative
geometry is incorporated into the same gauge-fixed variational structure
as the Onsager dissipative sector.

\subsection{The localized BF-type expression}

The localized BF-type expression should therefore not be interpreted
as a second independent dynamical principle or as a complete BF theory.
Rather, after the Nambu-adapted gauge condition is imposed, it provides
a differential-form representation of the coupling $A_{i}v_{H}^{i}$
on the extended thermodynamic state space.

\begin{equation}
\left.I_{\mathrm{pre}}\right|_{\chi_{H}=0}=I_{\mathrm{NNET}}^{\mathrm{loc}}.
\end{equation}

\subsection{Interpretation of $S_{\mathrm{NB}}$}

The scalar $S_{\mathrm{NB}}$ should not in general be identified with
the physical thermodynamic entropy. In the potential sector it generates
the dissipative drift through

\begin{equation}
A_{i}=\partial_{i}S_{\mathrm{NB}}.
\end{equation}

Its nonnegative quadratic production rate

\begin{equation}
\sigma_{\mathrm{NB}}=L^{ij}\partial_{i}S_{\mathrm{NB}}\partial_{j}S_{\mathrm{NB}}
\end{equation}

characterizes the dissipative sector of the present theory. Its identification
with physical entropy production requires additional thermodynamic
input.

\subsection{BRST structure and finite residual transformations}

The BRST structure developed in Sec.~5 is a consequence of the previously
specified gauge redundancy and gauge-fixing conditions, not an independent
justification for them. In the Nambu-adapted gauge, the corresponding
Faddeev--Popov operator becomes, on shell, the negative of the material
derivative along the full NNET flow. Its time-independent zero-mode
profiles therefore
satisfy the same equation as first integrals of the deterministic
dynamics. This is the linearization of the finite residual
Hamilton--Jacobi equation~\eqref{eq:finite_residual_hj}. At finite
amplitude, the residual parameter is nonincreasing along the original
NNET flow, with its rate fixed by the Onsager quadratic form as in
Eq.~\eqref{eq:residual_monotonicity}. In a drift-free region with
constant $L^{ij}$, its spatial gradient also gives the inviscid
Burgers reduction in Eq.~\eqref{eq:residual_burgers}.

\subsection{The new content of the present formulation}

The new content of the present formulation is therefore not the NNET
evolution equation by itself. Rather, it lies in the common gauge-fixed
origin of the Onsager and NNET constitutive relations, the identification
of the Nambu characteristic direction as the direction entering the
generalized gauge condition, the equivalence between the localized
BF-type representation and the Nambu-adapted gauge-fixed worldline
action, and the resulting Faddeev--Popov operator along the full
NNET flow. The finite residual Hamilton--Jacobi equation extends this
last structure beyond its infinitesimal limit and links residual
changes of the constitutive drift to the same Onsager tensor.

\subsection{Geometric hierarchy}

The integrable potential sector, Onsager gauge fixing, and Nambu-adapted
gauge fixing are summarized in Eq.~\eqref{eq:hierarchy}.

\subsection{Future directions}

A first important direction concerns the global structure of $S_{\mathrm{NB}}$.
In the present formulation, $S_{\mathrm{NB}}$ is introduced locally
as a dissipation potential. When a single global potential does not
exist, the theory should instead be formulated in terms of local potentials
defined on overlapping patches, together with appropriate transition
data specifying how these local descriptions are glued together.

\LyXZeroWidthSpace{}

A second, distinct direction concerns genuinely non-integrable thermodynamic
connections, for which $F=d_{a}A\neq0$. Unlike a closed but globally
non-exact connection, such a connection cannot be written as $A=d_{a}S_{\mathrm{NB}}$
even locally in a neighborhood where $F\neq0$.

As discussed in the original gauge-fixing formulation, such a non-vanishing
curvature corresponds to a situation in which the thermodynamic force
cannot be represented globally as the gradient of a single scalar
potential. Extending the present NNET construction to this non-integrable
sector may provide a natural framework for describing genuinely path-dependent
thermodynamic processes and more general nonequilibrium cycles.

\LyXZeroWidthSpace{}

Another natural direction is the stochastic extension of the present
theory through the Onsager--Machlup--Hashitsume path-integral formulation.
Such an extension would make it possible to study fluctuations around
the deterministic NNET trajectories and to clarify more systematically
the relation among the dissipation potential $S_{\mathrm{NB}}$, stochastic
path weights, and nonequilibrium stationary distributions.

\LyXZeroWidthSpace{}

Future directions also include the quantization of the thermodynamic
gauge field, the exploration of its connection with fluctuation theorems~\cite{Crooks_1999,Seifert_2012}, and possible links to information
geometry and contact geometry, which may further clarify the geometric
structure of nonequilibrium processes. In particular, contact geometry~\cite{Bravetti_2017} provides a related geometric framework based
on the contact form $\eta=dS-A_{i}da^{i}$, its exterior derivative
$d\eta$, and the Reeb vector field $\xi=\partial_{S}$. A contact
Hamiltonian $\mathcal{H}_{s}(S,A,a)$ then specifies the dynamics.
The relation between volume contraction in contact dynamics and volume-changing
structures in Nambu nonequilibrium thermodynamics may therefore be
worth investigating in greater detail. A precise formulation of this
relation is left for future work.

\LyXZeroWidthSpace{}

The present paper focuses on the geometric and variational foundations
of Nambu nonequilibrium thermodynamics. Applications to explicit physical
systems are discussed elsewhere~\cite{katagiri2025nambu3}.

\section*{Acknowledgments}

The author would like to express sincere gratitude to \textbf{Akio
Sugamoto} and \textbf{Tatsuaki Wada} for their valuable comments and
discussions. Special thanks are also due to \textbf{Shiro Komata}
for carefully reading the manuscript and providing numerous insightful
suggestions. The author further thanks \textbf{Ken Yokoyama} for sharing
several helpful perspectives on the gauge-theoretic formulation developed
in this work.

\section*{Declaration of generative AI and AI-assisted technologies in the manuscript preparation process}

During manuscript preparation, the author used ChatGPT (OpenAI)
for assistance with English-language revision, organization and
clarification of the exposition, checks of notation and algebraic
consistency, and LaTeX preparation. The author reviewed the resulting
material, made the necessary revisions, and accepts full responsibility
for the scientific content and conclusions of this article.

\bibliographystyle{unsrt}
\bibliography{nnet_new_typography_final}

\end{document}